\documentclass[%
reprint, 
 aps,pre,
 amsmath,amssymb,
]{revtex4-1}

\usepackage{amsmath}
\usepackage{xcolor}
\colorlet{mylinkcolor}{blue!66!black!80}
\usepackage[colorlinks=true,linkcolor=mylinkcolor,citecolor=mylinkcolor,filecolor=cyan,urlcolor=mylinkcolor,breaklinks=true]{hyperref}
\usepackage{mathtools}
\usepackage[capitalize]{cleveref}
\usepackage{braket}
\usepackage{bbold}

\newcommand{\avg}[1]{\langle#1\rangle}

\newcommand{\e}{{\rm e}}

\usepackage[version=4]{mhchem}

\begin{document}

\setcounter{page}{1} 

\title{Exponential volume dependence of entropy-current fluctuations at first-order phase transitions in chemical reaction networks}
\author{Basile Nguyen}
\author{Udo Seifert}
\email[For correspondence: ]{useifert@theo2.physik.uni-stuttgart.de}
\affiliation{II. Institut f{\"u}r Theoretische Physik, Universit{\"a}t Stuttgart, 70550 Stuttgart, Germany}
\date{\today}%

\begin{abstract}
{
In chemical reaction networks, bistability can only occur far from equilibrium. It is associated with a first-order phase transition where the control parameter is the thermodynamic force. At the bistable point, the entropy production is known to be discontinuous with respect to the thermodynamic force. We show that the fluctuations of the entropy production have an exponential volume-dependence when the system is bistable. At the phase transition, the exponential prefactor is the height of the effective potential barrier between the two fixed-points. Our results obtained for Schl{\"o}gl's model can be extended to any chemical network.
}
{}{}
\end{abstract}

\maketitle 

\section{Introduction}

Nonequilibrium phase transitions have been long studied and still remain less understood than their equilibrium counterparts. Most biological systems operate far from equilibrium and can achieve rich dynamics such as biochemical switching and oscillations, which are both observed, for example, in interlinked GTPases \cite{mizu12,suda13,beme15,ehrm19} or in the MinDE system \cite{fisc10,hala12,xion15,wu16,denk18}. Their complex behavior can be understood with simple chemical networks introduced by Schl{\"o}gl in the seventies to study nonequilibrium first- and second-order phase transitions \cite{schl72}. Subsequently, order parameters with their associated variances have been derived for these systems \cite{jans74,mcne74,math75,nico77,nico78,nico86}. 

In this paper, we focus on biochemical switches that undergo a first-order phase transition upon activation. From a deterministic perspective, this phase transition is associated with bistability where two stable steady states can coexist. In contrast, in the stochastic perspective, the steady-state is unique and is associated with a bimodal density distribution \cite{vell07,vell09}. In the thermodynamic limit, the stochastic system will relax to the more stable fixed-point except at the bistable point \cite{ge09,ge11}. There are two timescales relevant for a biochemical switch: a fast relaxation to the nearest fixed-point and a slower transition between the states, where the coarse-grained transition rates are proportional to the exponential of the inverse volume \cite{hang84,hinc05}.  In chemical reaction networks, bistability can occur only far from 
equilibrium since an equilibrium distribution will always be a Poisson distribution distribution, and, thus, have a single peak 
\cite{heue06,gard04}.

The behavior of the entropy production at first- and second-order phase transitions has been investigated in many systems such as chemical networks or nonequilibrium Ising models \cite{xiao08,ge09,vell09,croc05,andr10,ge11,rao11,bara12,tom12,zhan16,fala18,nguy18,noa19}. At first-order phase transitions, the entropy production rate has a discontinuity with respect to the thermodynamic force whereas at second-order phase transitions its first derivative has a discontinuity. Recently, it has been shown that the critical fluctuations of the entropy production diverge with a power-law with the volume at a second-order phase transition \cite{nguy18}. For first-order phase transitions, the behavior of entropy production fluctuations has not been investigated yet to the best of our knowledge.

We will show that the fluctuations of the entropy production have an exponential volume-dependence at first-order phase transitions in chemical reaction networks. Our results are obtained for Schl{\"o}gl's model. First, we compute the entropy fluctuations numerically from the chemical master equation using standard large deviation techniques \cite{koza99,touc09}. Second, we compute the current fluctuations for a coarse-grained two-state model and show that the diffusion coefficient diverges at the bistable point with an exponential prefactor given by the height of the effective potential barrier separating the two fixed-points.  

The paper is organized as follows. In \cref{sec:mod_def}, we introduce Schl{\"o}gl's model and define the entropy production. In \cref{sec:CME}, we consider the chemical master equation and compute the diffusion coefficient numerically. In \cref{sec:twostate}, we introduce an effective two-state model and compute an analytical expression for the diffusion coefficient. We conclude in \cref{sec:conclusion}.

\section{Schl{\"o}gl's model and entropy production} \label{sec:mod_def}
\subsection{Model definition}
The Schl{\"o}gl model is a paradigmatic model for biochemical switches \cite{schl72,vell09}. It consists of a chemical species $X$ in a volume $\Omega$. The external bath contains two chemical species $A$ and $B$ at fixed concentrations $a$ and $b$, respectively. The set of chemical reactions is
\begin{equation}
\begin{split}
  2X+A &\xrightleftharpoons[{k_{-1}}]{k_1} 3X,\\
  B &\xrightleftharpoons[{k_{-2}}]{k_2} X,
\end{split}
\label{eq:schlogl_reac}
\end{equation}
where $k_1, k_{-1}, k_{2}$ and $k_{-2}$ are transition rates. The system is driven out of equilibrium due to a difference of chemical potential between $A$ and $B$, which is written as  $\varDelta\mu \equiv \mu_A-\mu_B$. A cycle in which an $X$ molecule is created with rate $k_1$, and then degraded with rate $k_{-2}$ leads to the consumption of a substrate $A$ and generation of a product $B$. The thermodynamic force associated with this cycle is  
 \begin{equation} 
 {\varDelta\mu} \equiv \ln \frac{k_{-2} k_1a}{k_{-1}k_2b},
 \label{eq:schlogl_ltb}
\end{equation} 
where the temperature $T$ and Boltzmann's constant $k_B$ are set to $1$ throughout this paper. The above relation between the thermodynamic force $\varDelta\mu$ and the transition rates is known as generalized detailed balance.

\subsection{Entropy production}

Along a stochastic trajectory $n(t)$, where $n$ labels the state with $n$ molecules of species $X$, 
the entropy production change of the medium can be identified as \cite{seif12}
\begin{equation}
 \varDelta s^m = Z_B(t) \ln \frac{k_{-2}}{k_2b} + Z_A(t)\ln \frac{k_1a}{k_{-1}}.
 \label{eq:deltas_stoch}
\end{equation}
 Here, $Z_A(t)$ and $Z_B(t)$ are random variables which count transitions in the $A$ and $B$ channel, respectively. For example, $Z_B(t)$ increases by one if a $B$ is produced, which happens if a reaction with rate $k_{-2}$ takes place. Likewise, it decreases by one if a $B$ is consumed, which happens if a reaction with rate $k_2$ takes place, i.e.
\begin{equation}
\begin{aligned}
\ce{$2X\,+\,A$ &->[$k_{1}$]$3X$}\qquad &Z_A\rightarrow Z_A+1 \, , \\
\ce{$3X$ &->[$k_{-1}$]$2X\,+\,A$}\qquad &Z_A\rightarrow Z_A-1 \, , \\
\ce{$X$ &->[$k_{-2}$]$B$}\qquad &Z_B\rightarrow Z_B+1 \, , \\
\ce{$B$ &->[$k_2$]$X$}\qquad &Z_B\rightarrow Z_B-1 \, .
\end{aligned}
\label{eq:zb_def}
\end{equation}
In this paper, we focus on $\varDelta s^m$, the extensive part of the total entropy production $\varDelta s^\textrm{tot}\equiv \varDelta s^m + \varDelta s$. The remaining part is the change in stochastic entropy \cite{seif05}
\begin{equation}
 \varDelta s = -\ln p_{n_t}(t) + \ln p_{n_0}(0),
\end{equation}
where $p_{n_t}(t)$ is the probability to find the system in state $n_t$ at time $t$.

Using \cref{eq:deltas_stoch,eq:zb_def}, we can write the mean entropy production rate per volume in the steady-state as \cite{seif12}
\begin{equation}
\begin{aligned}
 \sigma &\equiv \lim_{t\to\infty} \frac{\avg{\varDelta s^m}}{t\Omega} \\
  &=\lim_{t\to\infty} \left[ \frac{\avg{Z_B(t)}}{t\Omega} \ln \frac{k_{-2}}{k_2b} +  \frac{\avg{Z_A(t)}}{t\Omega}\ln \frac{k_1a}{k_{-1}}\right]  \\
  & = J_B \varDelta\mu.
 \label{eq:schlogl_sigma}
 \end{aligned}
\end{equation}
In the steady-state, the mean flux density of $B$ molecules 
\begin{equation}
 J_B \equiv \lim_{t\to\infty} \frac{\avg{Z_B(t)}}{t\Omega} = \lim_{t\to\infty} \frac{\avg{Z_A(t)}}{t\Omega}
 \label{eq:schlogl_Jnum}
\end{equation}
is equal to the flux of $A$ molecules consumed.

We can quantify the fluctuations of $B$ molecules with the diffusion coefficient
\begin{equation}
\begin{aligned}
 D_B &\equiv \lim_{t\rightarrow\infty} \frac{\avg{Z_B(t)^2}-\avg{Z_B(t)}^2}{2 t \Omega} \\
 &= D_A \equiv \lim_{t\rightarrow\infty} \frac{\avg{Z_A(t)^2}-\avg{Z_A(t)}^2}{2 t \Omega},
 \label{eq:schlogl_DBnum}
 \end{aligned}
\end{equation}
where we prove the second equality, $D_B=D_A$, in \cref{app:cumul}.  Specifically, we show there that $Z_A(t)$ and $Z_B(t)$ have the same cumulants. 

Finally, using \cref{eq:deltas_stoch,eq:schlogl_DBnum} we obtain the diffusion coefficient associated with the entropy production in the medium as 
\vspace{1cm}
\begin{widetext}
\begin{equation}
\begin{aligned}
D_\sigma&\equiv \lim_{t\rightarrow\infty} \frac{\avg{(\varDelta s^\textrm{bath})^2}-\avg{\varDelta s^\textrm{bath}}^2}{2 t \Omega} \\
&= D_B \left(\ln\frac{k_{-2}}{k_2b}\right)^2 +D_A \left(\ln\frac{k_1a}{k_{-1}}\right)^2 + \lim_{t\rightarrow\infty} \left[\frac{\avg{Z_A(t)Z_B(t)} - \avg{Z_A(t)}\avg{Z_B(t)}}{2t\Omega} \right] 2\left(\ln\frac{k_{-2}}{k_2b}\right)\left(\ln\frac{k_1a}{k_{-1}}\right) \\
&=  D_B \varDelta\mu^2.
 \label{eq:schlogl_Dsigma}
  \end{aligned}
\end{equation}
\end{widetext}
Since the stochastic entropy production is not extensive in time, this diffusion coefficient is equal to the one for the total entropy production. 

\section{Chemical master equation} \label{sec:CME}
\subsection{Stationary solution}
The state of the system is fully determined by the total number $n$ of $X$ molecules. The time evolution of $P(n,t)$, which is the probability to find the system in state $n$ at time $t$, is governed by the chemical master equation (CME) 
\begin{equation}
\begin{aligned}
 \partial_t P(n,t) &= f_{n-1}P(n-1,t) + g_{n+1}P(n+1,t) \\
 &- (f_n+g_n)P(n,t).
 \label{eq:schlogl_CME}
 \end{aligned}
\end{equation}
Here, we define the rate parameters as
\begin{equation}
\begin{aligned}
 f_n = \alpha_n^+ + \beta_n^+ &\equiv \frac{ak_1 n(n-1)}{\Omega} + bk_2 \Omega, \\
 g_n = \alpha_n^- + \beta_n^- &\equiv \frac{k_{-1} n(n-1)(n-2)}{\Omega^2} + k_{-2} n.
 \label{eq:ratepara_one}
\end{aligned}
\end{equation}
The system can reach a nonequilibrium steady-state with a distribution written as $P_n$. The analytical solution for this steady-state probability distribution reads
\begin{equation}
\begin{aligned}
 \frac{P_n}{P_0} &= \prod_{i=0}^{n-1} \frac{f_i}{g_{i+1}}, \\
 = &\frac{f_0}{g_1} \sqrt{\frac{f_1 g_1}{f_n g_n}} \\
 &\exp \left[ \frac{1}{2}\ln\left( \frac{f_1}{g_1}\right) + \sum_{i=2}^{n-1}\ln\left(\frac{f_i}{g_i}\right) +\frac{1}{2}\ln\left(\frac{f_n}{g_n}\right) \right] \\
 & (n\geq3),
 \label{eq:prob_one}
 \end{aligned}
\end{equation}
where the normalization is given by 
\begin{equation}
P_0=1-\sum_{j=1}^\infty P_j.
\end{equation}

For a large number of states ($\Omega\to\infty$), we can write $x=n/\Omega$ as a continuous variable and approximate the exponential by an integral using the trapezium rule, which is valid if ${f_n}/{g_n}$ is bounded. Without loss of generality, we will choose parameters such that the fixed-points are far enough from the boundary, which means that the probability to find the system close to $x=0$ is negligible. Note that the case where the solution does not vanish at the boundary is discussed in details in \cite{hinc05}. 

From \cref{eq:prob_one}, the continuous steady-state distribution can be written as
\begin{equation}
 p(x) \propto \exp\left[-\Omega \left(\phi_0(x)+ \frac{1}{\Omega}\phi_1(x)\right)\right]
 \label{eq:schlogl_px}
\end{equation}
where we define the non-equilibrium potential 
\begin{equation}
 \phi_0(x)\equiv-\int_0^x \textrm{d}y\, \ln\left(\frac{f(y)}{g(y)}\right),
 \label{eq:schlogl_phi0}
\end{equation}
and
\begin{equation}
 \phi_1(x)\equiv -\frac{1}{2} \ln \frac{1}{f(x)g(x)}.
 \label{eq:schlogl_phi1}
\end{equation}
Here, we have defined the total transition rates 
\begin{equation}
\begin{aligned}
 f(x) &\equiv \alpha^+(x) + \beta^+(x) =  a k_1 x^2 + k_2 b,\\
 g(x) &\equiv \alpha^-(x) + \beta^-(x) =  k_{-1} x^3 + k_{-2} x.
 \label{eq:schlogl_fg}
\end{aligned}
\end{equation}

In the deterministic limit ($\Omega\to\infty$), we obtain the equation of the time evolution of the density,
\begin{equation}
 \bar{x} \equiv \sum_{n} n P(n,t)/\Omega
\end{equation}
 as 
\begin{equation}
 \frac{\mathrm{d}\bar{x}}{\mathrm{d}t} = f(\bar{x})-g(\bar{x})
 \label{eq:schlogl_Det}
\end{equation}
from the chemical master \cref{eq:schlogl_CME}. In the steady-state, this equation has three solutions $(x_-,x_0,x_+)$. Bistability occurs when all solutions are real, we order the fixed-points as follows: $0<x_-<x_0<x_+$, where $x_\pm$ are stable (i.e. $f'(x_\pm)<g'(x_\pm)$) and $x_0$ is unstable (i.e. $f'(x_0)>g'(x_0)$). 

\subsection{Behavior of the entropy production at the phase transition} \label{sec:res}
The mean entropy production rate, defined in \cref{eq:schlogl_sigma}, can be written using \cref{eq:schlogl_fg} as 
\begin{equation}
\begin{aligned}
 \sigma &= \varDelta\mu\int_0^\infty \mathrm{d}x \left[\beta^-(x)-\beta^+(x)\right]p(x), \\
  \label{schlogl_J_Bna}
\end{aligned}
\end{equation}
In the thermodynamic limit, the stochastic system will relax to the more stable fixed-point except at the bistable point \cite{ge09,ge11}. Consequently, the rate of entropy production will be discontinuous with respect to the thermodynamic force at the bistable point. Specifically, with increasing $\varDelta\mu$, $\sigma$ will jump from $\left[\beta^-(x_-)-\beta^+(x_-)\right]\varDelta\mu $ to $\left[\beta^-(x_+)-\beta^+(x_+)\right]\varDelta\mu$, which are the rates of entropy production at the two fixed-points $x_-$ and $x_+$, respectively.

We now derive an expression for the fluctuations of the entropy production. We follow an approach based on large deviation theory \cite{koza99,touc09,touc18} and considered for Brownian ratchets in \cite{uhl18}. We want to compute the cumulants related to the number of produced $B$ molecules. They are obtained through the scaled cumulant generating function (SCGF)
\begin{equation}
 \alpha(\lambda) \equiv \lim_{t\rightarrow\infty} \frac{1}{t} \ln \avg{\e^{\lambda Z_B}}
 \label{eq:SCGF}
\end{equation}
where $Z_B$ is the time-integrated current of $B$ molecules defined in \cref{eq:zb_def}. Note that $\alpha(\lambda)$ is unrelated to the transition rate defined in \cref{eq:ratepara_one}. From now on, we will drop the $t$ dependence on $Z_B$ for readability. Expanding the generating function yields 
\begin{equation}
 \alpha(\lambda) = \Omega J_B \lambda + \Omega D_B \lambda^2 + \mathcal{O}(\lambda^3)
   \label{eq:alphaexp}
\end{equation}

The SCGF can be obtained by considering the moment generating function 
\begin{equation}
\begin{aligned}
 g(\lambda,t) &\equiv \avg{\lambda \e^{\lambda Z_B}}  \\ 
 &= \sum_n g(\lambda,n,t)P(n,t) 
 \end{aligned}
\end{equation}
where we define
\begin{equation}
\begin{aligned}
 g(\lambda,n,t) &\equiv \avg{\e^{\lambda Z_B}|n(t)=n} \\
 &= \sum_{Z_B} e^{\lambda Z_B} P(n,Z_B,t),
\end{aligned}
\label{eq:gdef}
 \end{equation}
which is conditioned on the final state of the trajectory $n(t)$. The time evolution of this quantity is given by 
\begin{equation}
 \partial_tg(\lambda,n,t) = \sum_m \mathcal{L}_{nm}(\lambda)g(\lambda,m,t).
 \label{eq:timeevolgn}
\end{equation}
where $\mathcal{L}(\lambda)$ is the tilted operator. Note that for $\lambda=0$, it is identical to the operator generating the time evolution of the probability distribution in \cref{eq:schlogl_CME} 

We want to specify $\mathcal{L}(\lambda)$ for the chemical master \cref{eq:schlogl_CME}. First, we write the time evolution of the probability distribution as
\begin{equation}
  P(n,Z_B,t) =  \sum_m w_{mn}P(m,Z_B-d_{mn},t) - w_{nm}P(n,Z_B,t)
  \label{eq:timevolprob}
\end{equation}
where $w_{mn}$ is the transition rate from state $m$ to state $n$ and $d_{mn}$ is the distance matrix which characterize how $Z_B$ changes during a transition. In the case of the CME, \cref{eq:timevolprob} reduces to  
\begin{equation}
\begin{aligned}
  P(n,Z_B,t) &= \alpha_{n+1}^-P(n+1,Z_B,t) \\
             &+ \alpha_{n-1}^-P(n-1,Z_B,t)\\
             &+ \beta_{n+1}^-P(n+1,Z_B+1,t)  \\
             &+ \beta_{n-1}^+P(n-1,Z_B-1,t) \\
             &- \left(\alpha_n^++\beta_n^++\alpha_n^-+\beta_n^-\right)P(n,Z_B,t)
  \label{eq:timevolprobCME}
  \end{aligned}
\end{equation}
where the rates $\alpha_n^\pm$ and $\beta_n^\pm$ are given by \cref{eq:ratepara_one}. Using \cref{eq:gdef}, we can write 
\begin{equation}
\begin{aligned}
 \partial_t g(\lambda,n,t) &= \sum_m  w_{mn}\sum_{Z_B} \e^{\lambda Z_B} P(m,Z_B-d_{mn},t)\\
 &- w_{nm} g(\lambda,n,t).
  \end{aligned}
  \end{equation}
  With a change of variable, we obtain
  \begin{equation}
\begin{aligned}
 \partial_t g(\lambda,n,t) &=  \sum_m w_{mn} \sum_{Y_B} \e^{\lambda(Y_B+d_{mn})} P(m,Y_B,t)\\
 &- w_{nm} g(\lambda,n,t) \\
 &= \sum_m\e^{\lambda d_{mn}}w_{mn} g(\lambda,m,t) - w_{nm}g(\lambda,n,t), \\
 \end{aligned}
 \label{eq:dgdt}
\end{equation}
which specifies the tilted operator $\mathcal{L}(\lambda)$ defined in \cref{eq:timeevolgn}.  

The cumulants can be obtained by solving the eigenvalue equation $\mathcal{L}(\lambda) \mathbf{Q}(\lambda) = \alpha(\lambda) \mathbf{Q}(\lambda)$, where 
\begin{equation}
 \mathcal{L}(\lambda) = \mathcal{L}_0 +  \mathcal{L}_1 \lambda +  \mathcal{L}_2\lambda^2 + \mathcal{O}(\lambda^3)
\end{equation}
and the distribution
\begin{equation}
 \mathbf{Q}(\lambda)  = \mathbf{P} + \mathbf{Q}_1\lambda +   \mathbf{Q}_2\lambda^2 + \mathcal{O}(\lambda^3).
\end{equation}
Sorting by orders of $\lambda$, we obtain
\begin{equation}
\begin{aligned}
 \mathcal{L}_0 \mathbf{P} &= 0,   \\ 
 \mathcal{L}_0 \mathbf{Q}_1 + \mathcal{L}_1 \mathbf{P} &= \Omega J_B \mathbf{P}, \\
 \mathcal{L}_1 \mathbf{Q}_1 + \mathcal{L}_0 \mathbf{Q}_2 +  \mathcal{L}_2 \mathbf{P} &= \Omega D_B \mathbf{P} + \Omega J_B \mathbf{Q}_1. 
\end{aligned} \label{eq:Ltilde}
\end{equation}
We multiply these equations with $\bra{1}$ on the left-hand side and note that $\mathbf{P}$ is normalized, i.e. $\Braket{1|\mathbf{P}}=0$, where $\braket{\cdot | \cdot}$ denotes the standard scalar product. We can compute the mean flux density
\begin{equation}
 J_B = \frac{1}{\Omega} \bra{1}\mathcal{L}_1 \ket{\mathbf{P}} = \frac{1}{\Omega}\sum_n \left(\beta_n^- - \beta_n^+\right)P_n
\end{equation}
 and the diffusion coefficient 
\begin{equation}
\begin{aligned}
 D_B &= \frac{1}{\Omega} \Big(\bra{1}\mathcal{L}_1  \ket{\mathbf{Q}_1} + \bra{1}\mathcal{L}_2  \ket{\mathbf{P}}\Big) - J_B \Braket{1 | \mathbf{Q}_1}\\
 &= \frac{1}{\Omega} \sum_n\left(\left(\beta_n^--\beta_n^+\right)(Q_1)_n + \frac{1}{2}\left(\beta_n^-+\beta_n^+\right)P_n\right)\\
 &- J_B \sum_n (Q_1)_n.
 \label{eq:DCME}
 \end{aligned}
\end{equation}
The mean rate of entropy production $\sigma$ as well as its associated diffusion coefficient $D_\sigma$ can be evaluated using \cref{eq:schlogl_sigma,eq:schlogl_Dsigma}.

\subsection{Numerical results} \label{sec:CME_numres}

 Throughout this paper, we set the parameters to $k_1 = 1, k_2 = 0.2, k_{-2} = 1, a = 1$ and $b = 1$. The transition rate $k_{-1}$ is computed from $\varDelta\mu$ and the generalized detailed balance relation \cref{eq:schlogl_ltb}, where $\varDelta\mu$ is a control parameter of the phase transition.  
 
\begin{figure}
 \includegraphics[width=0.9\linewidth]{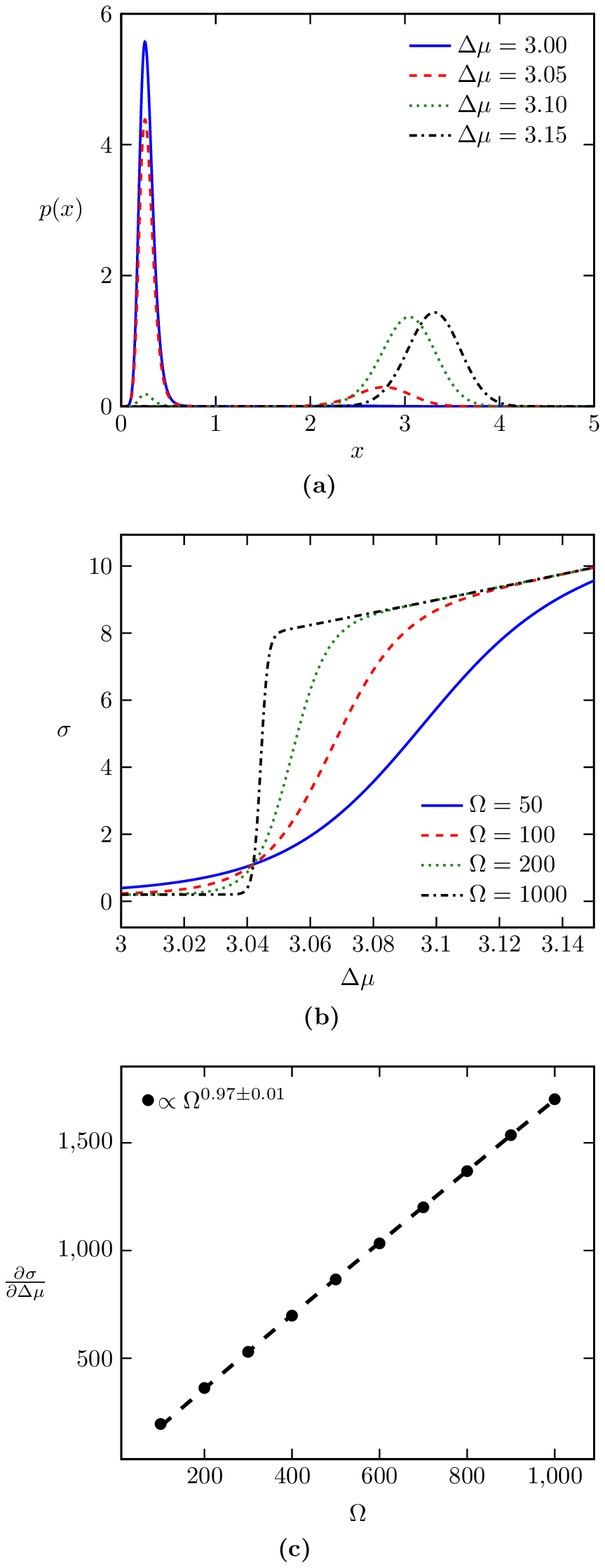}
 \caption{\label{fig:pss}  Phase transition in the Schl{\"o}gl model. (a) Stationary distribution of chemical species $X$ for $\Omega=100$ and different values of $\varDelta\mu$. (b) Mean entropy production rate $\sigma$ as a function of $\varDelta\mu$ for different system sizes $\Omega$.  (c) Maximum of the first derivative of $\sigma$ as a function of the system size $\Omega$. Parameters are given in the main text.}
\end{figure}

 In \cref{fig:pss}(a), we plot the stationary distribution $p(x)$ which is bimodal in the vicinity of the phase transition ($\varDelta\mu^\textrm{bi} \simeq 3.045$). In \cref{fig:pss}(b), we plot the entropy production rate $\sigma$ as a function of $\varDelta\mu$. With increasing system size, $\sigma$ gets steeper at the bistable point. In \cref{fig:pss}(c), we show that the first derivative of $\sigma$ follows a power-law with an effective prefactor close to $1$. In the thermodynamic limit ($\Omega\rightarrow\infty$), the entropy production rate becomes discontinuous as shown by Ge and Qian \cite{ge09,ge11}. 

\begin{figure}
 \includegraphics[width=0.9\linewidth]{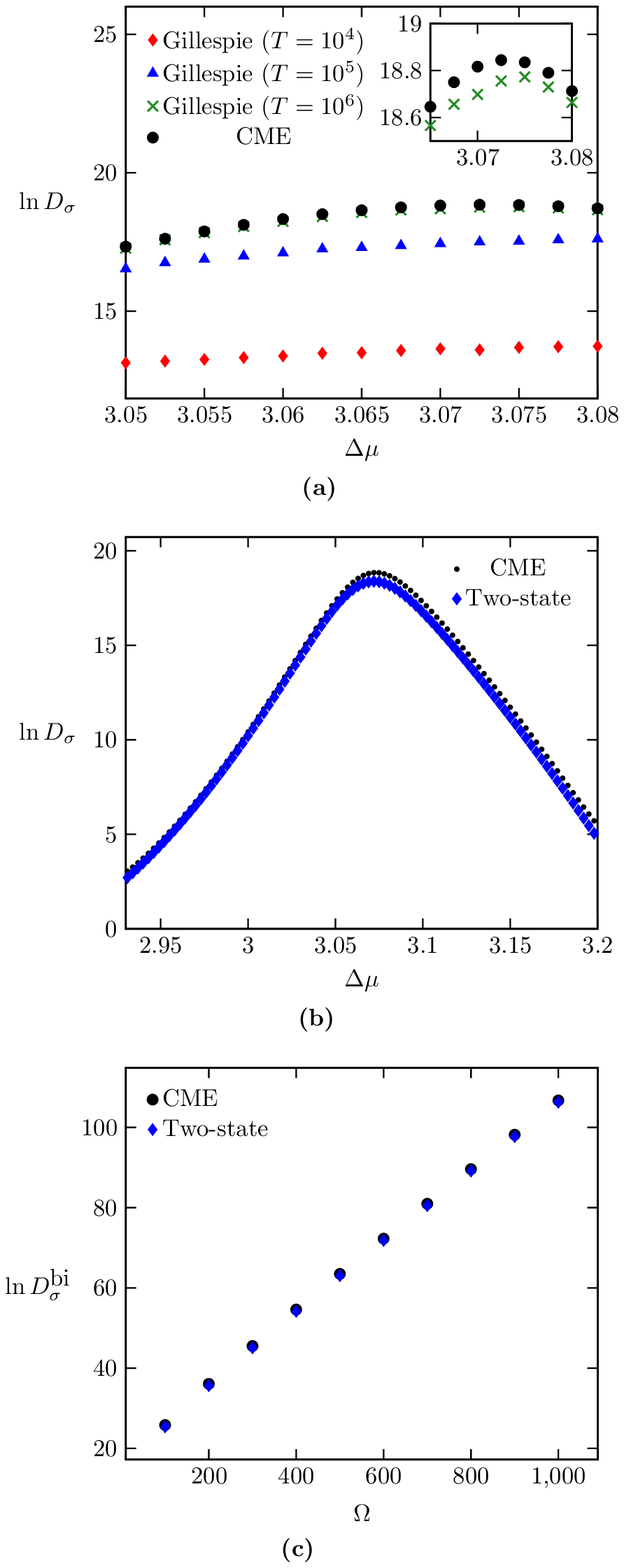}
 \caption{\label{fig:diff}  Behavior of the diffusion coefficient $D_\sigma$ close to the bistable point. (a) Diffusion coefficient from simulations using Gillespie's algorithm for $10^4$ trajectories with a sampling time of $T=10^4, 10^5, 10^6$. For the CME, we solve the linear system given by \cref{eq:DCME} numerically for $\Omega=100$. (b) Diffusion coefficient for the two-state model obtained by evaluating \cref{eq:finitediffscaling}. (c) Finite-size scaling of the maximum of the diffusion coefficients $D_\sigma^\text{bi}$. The corresponding slopes are given in \cref{tab:critexp}. }
\end{figure}

The diffusion coefficient $D_\sigma$ reaches a maximum at the bistable point and has an exponential volume-dependence. In \cref{fig:diff}(a), we compare simulations of the chemical master \cref{eq:schlogl_CME}  using Gillespie's algorithm \cite{gill77} for three increasing sampling times $T$ with the numerical solution obtained by solving the linear system given by \cref{eq:DCME}. The systematic difference between these two methods is due to a limited sampling time. In the next section, we present an effective two-state model and derive an analytical expression for the diffusion coefficient. 

\section{Two-state model} \label{sec:twostate}
\subsection{Stationary solution}
In the bistable regime, the system has two timescales. First, it will relax towards the nearest stable fixed-points $x_\pm$ and fluctuate around it. Close to the fixed-points, the system can be modeled by stationary Gaussian processes. Specifically, the distribution $p(x)$ can be expanded around its stable fixed-points $x_\pm$ as \cite{vell09},
\begin{equation}
 p(x) \approx \sum_{x_*=(x_-,x_+)}\frac{\e^{-\Omega \phi_0(x_*)}}{\mathcal{Z}_GA_2(x_*)} \exp \left[\frac{-\Omega \phi_0''(x_*) \left(x-x_* \right)^2}{2} \right],
 \label{eq:schlogl_pSN}
\end{equation}
where 
\begin{equation}
\begin{aligned}
 \mathcal{Z}_G &\equiv \sum_{x_*=(x_-,x_+)} \frac{\e^{-\Omega  \phi_0(x_*)}\sqrt{2\pi}}{A_2(x_*)\sqrt{\left|\phi_0''(x_*)\right|\Omega }}.\\
 \label{schlogl_JSN_ZG}
 \end{aligned}
\end{equation}
 
 Second, as stochastic fluctuations are always present, the system will at some point in time reach the unstable fixed-point $x_0$ beyond which it can relax towards the other fixed-point. Based on this behavior, the infinite-state system \cref{eq:schlogl_CME} can be coarse-grained into a two-state process between the stable fixed-points $x_\pm$ \cite{keiz79}. The transition rates from $x_\pm$ to $x_\mp$ depend exponentially on the system size and are given explicitly by \cite{hang84,hinc05} 
\begin{equation}
 r_\pm = \frac{\e^{-\Omega\left[\phi_0(x_0)-\phi_0(x_\pm)\right]}f(x_\pm)\sqrt{-\phi_0''(x_0)\phi_0''(x_\pm)}}{2\pi \Omega}.
 \label{eq:rpm}
\end{equation}

\subsection{Behavior of the entropy production at the phase transition} \label{sec:twostate_sigma}

We want to characterize the fluctuations of the entropy production for the two-state model. We consider the thermodynamic flux $J_B$ and its associated diffusion coefficient $D_B$ defined in \cref{eq:schlogl_Jnum,eq:schlogl_DBnum}. There are two contributions to both of these quantities. First, the system can fluctuate around one of the fixed-point, which is modeled by a Gaussian process \cref{eq:schlogl_pSN}. Within the state $x_\pm$, the thermodynamic flux $J_{\pm}$ and its diffusion coefficient $D_{\pm}$ can be computed exactly \cite{touc09}. Second, the system can jump from state $x_\pm$ to $x_\mp$ on the largest timescale.  For simplicity, we will assume that a transition from $x_\pm$ to $x_\mp$ produces an average flux of $\mathcal{B}_\pm$ molecules, where we expect $\mathcal{B}_\pm \sim \mathcal{O}(\Omega)$. The probability to be in state $x_\pm$ is 
\begin{equation}
  p_\pm \equiv \frac{r_\mp}{r_-+r_+}.
 \end{equation}
 We will compute $D_B$ using large deviation theory as introduced in \cref{sec:res}. In \cref{app:noLDT}, we compute $D_B$ without relying on large deviation theory.

Combining these two contributions, the tilted operator for this two-system system reads \cite{piet16}
 \begin{equation}
  \mathcal{L}(\lambda)=\begin{pmatrix}
- r_{-}+J_-\lambda+D_-\lambda^2 & r_+\e^{\lambda \mathcal{B}_+} \\
r_{-}\e^{\lambda \mathcal{B}_-} & -r_++J_+\lambda+D_+\lambda^2 \\
\end{pmatrix}.
\label{appeq:Ltilted}
\end{equation}
The maximal eigenvalue of $\mathcal{L}(\lambda)$ is 
\begin{equation}
 \alpha(\lambda) = \mathrm{Tr} \mathcal{L}(\lambda)/2 + \sqrt{\left( \mathrm{Tr} \mathcal{L}(\lambda) \right)^2/4 - \mathrm{Det}\mathcal{L}(\lambda)},
\end{equation}
where $\mathrm{Tr}$ and $\mathrm{Det}$ denote the trace and the determinant, respectively. From \cref{eq:alphaexp}, the average flux of $B$ is given by 
\begin{equation}
\begin{aligned}
 J_B &= \frac{\partial \alpha(\lambda)}{\partial \lambda}\Bigg|_{\lambda=0} \\
    &=  p_-J_- + p_+J_+ + r_-p_-(\mathcal{B}_-+\mathcal{B}_+) \\
     &=  p_-J_- + p_+J_+ + \mathcal{O}\left( \e^{-\Omega | \varDelta\phi|}\right)
 \end{aligned}
 \end{equation}
where $\varDelta\phi_0 = \phi_0(x_0)-\phi_0(x_-)$. The diffusion coefficient reads
\begin{equation}
\begin{aligned}
 D_B &=\frac{1}{2}\frac{\partial^2 \alpha(\lambda)}{\partial \lambda^2}\Bigg|_{\lambda=0} \\
 &= p_-p_+ \frac{\left(J_--J_+\right)^2}{r_-+r_+} + p_- D_- + p_+ D_+   \\
     &+p_-p_+\left(\mathcal{B}_-+\mathcal{B}_+\right)(p_+-p_-)(J_- -J_+) \\
     &+\frac{1}{2}\left(\mathcal{B}_-+\mathcal{B}_+\right)^2 p_-p_+ \left(r_-p_+ + p_-r_+ \right) \\ 
     &=p_-p_+ \frac{\left(J_--J_+\right)^2}{r_-+r_+} + \mathcal{O}(\Omega). \\     
 \label{eq:twostate_DB}
 \end{aligned}
\end{equation}

We insert the transition rates \cref{eq:rpm} into the previous expression and obtain
 \begin{equation}
   D_B\big|_{\varDelta\mu^\textrm{bi}} = p_-p_+ \left(J_--J_+\right)^2 \frac{\e^{\Omega\left[\phi_0(x_0)-\phi_0(x_\pm)\right]}\pi \Omega }{f(x_-)\sqrt{-\phi_0''(x_0)\phi_0''(x_\pm)}},
   \label{eq:finitediffscaling}
 \end{equation}
where $f(x_-)=f(x_+)$ and $\phi_0(x_-)=\phi_0(x_+)$ at the bistable point. As a main result, we have thus shown that the diffusion coefficient scales as 
\begin{equation}
 D_B\big|_{\varDelta\mu^\textrm{bi}} \propto \e^{\Omega\left[\phi_0(x_0)-\phi_0(x_\pm)\right]},
  \label{eq:infdiffscaling}
\end{equation}
where the exponential prefactor is the height of the effective potential barrier between the two fixed-points. The mean rate of entropy production $\sigma$ as well as its associated diffusion coefficient $D_\sigma$ can be evaluated using \cref{eq:schlogl_sigma,eq:schlogl_Dsigma}.

Here, we have assumed that the average flux of $B$ molecules $\mathcal{B}_\pm$ produced during a jump from $x_\pm$ to $x_\mp$ is known and inserted these values in \cref{appeq:Ltilted}. In fact, a trajectory from $x_-$ to $x_+$ will produce a path-dependent flux of $\mathcal{B}_\pm$. When we perform a coarse-graining of the CME into a two-state model, we lose this information. Nevertheless, as we are only interested in the leading terms of $D_B$, we have shown that the contributions from jumps between the fixed-points $\mathcal{B}_\pm$ can be neglected close to the bistable point for large system sizes.

\begin{table}
\begin{tabular}{|l|c|}
  \hline
  Method &  Exponential prefactor $\delta$ ($D_\sigma\propto \e^{\delta\Omega}$) \\ \hline
  \cref{eq:DCME} & $0.0846 \pm 0.0005$ \\ \hline
  \cref{eq:finitediffscaling} & $0.0846 \pm 0.0005$ \\ \hline
  \cref{eq:infdiffscaling} & $0.0823$ \\ \hline
 \end{tabular}
 \caption{\label{tab:critexp} Scaling of the diffusion coefficient $D_\sigma$ obtained with the CME, \cref{eq:DCME}, the two-state model, \cref{eq:finitediffscaling}, and by evaluating  the height of the effective potential barrier separating the two fixed-points, \cref{eq:infdiffscaling}. The maximum logarithm of the diffusion coefficient is fitted with an linear function of the system size $\Omega$ and the errors is given with 95\% confidence bounds.}
\end{table}

\subsection{Numerical results} \label{sec:twostate_numres}
We now compare the analytical results with numerical evaluations the CME. In \cref{fig:diff}(b), we show $D_\sigma$, \cref{eq:finitediffscaling}, and compare it with the numerical results from the CME. We find that $D_\sigma$ evaluated for the two-state model almost matches the CME close to the bistable point. In \cref{fig:diff}(c), we show that the diffusion coefficient has an exponential volume-dependence at the bistable point. In \cref{tab:critexp}, we compare the scaling of the maximum of the diffusion coefficient obtained numerically and by evaluating \cref{eq:infdiffscaling}. The difference between the numerical prefactors and our analytical expression, \cref{eq:finitediffscaling,eq:infdiffscaling}, is due to finite-size effects.


\section{Conclusion} \label{sec:conclusion}

We have investigated the fluctuations of the entropy production at the phase transition occurring in a paradigmatic model of biochemical switches. A control parameter for this phase transition is the thermodynamic force driving the system out of equilibrium. The mean entropy production rate has a discontinuity with respect to the thermodynamic force at the phase transition and fluctuations, which are quantified by the diffusion coefficient that diverges. First, we have computed the diffusion coefficient numerically for the chemical master equation. Second, we have derived an analytical expression of the diffusion coefficient for an effective two-state model. We find that the diffusion coefficient from the two-state model slightly underestimates the diffusion coefficient from the chemical master equation. This difference could be explained by the coarse-graining procedure, which is known to underestimate fluctuations far from equilibrium \cite{horo15}. Finally, we have shown that the diffusion coefficient has an exponential volume-dependence at the bistable point, where the exponential prefactor is given by the height of the effective potential barrier between the two fixed-points. 

In this paper, we have considered Schl{\"o}gl's model as a simple model for a nonequilibrium first-order phase transitions. We expect that models with additional chemical reactions or species show qualitatively the same behavior at the phase transition. For bistable systems with multiple species, one can introduce reaction coordinates along which the system becomes effectively one-dimensional. More generally, we expect that diffusion coefficients associated with currents or the entropy production can be computed at first-order phase transitions for a large class of nonequilibrium systems by describing them with discrete jump processes. The exponential volume dependence discussed here should then be generic for these cases

\section*{Acknowledgements}
We thank Matthias Uhl and Lukas P. Fischer for valuable discussions. 

\appendix

\section{Relation between the cumulants of currents in a system with two reaction channels}\label{app:cumul}
Here, we will prove that time-integrated currents $Z_A(t)$ and $Z_B(t)$, which are defined in \cref{eq:zb_def},  have the same cumulants. We will rely on the large deviation theory \cite{koza99,touc09,touc18}, which is introduced in \cref{sec:res}. 

The tilted operators are defined for general observables in \cref{eq:timeevolgn,eq:dgdt}. For the $A$ reaction channel, it reads 
\begin{equation}
\begin{aligned}
 (\mathcal{L}_A(\lambda))_{i,j} &\equiv \delta_{i,j+1}\left(\alpha_i^+\e^\lambda+\beta_i^+\right)+\delta_{i,j-1}\left(\alpha_i^-\e^{-\lambda}+\beta_i^-\right)\\
 &-\delta_{i,j} \left(\alpha_i^+ + \alpha_i^- + \beta_i^+ + \beta_i^+ \right) 
 \end{aligned}
\end{equation}
and for the $B$ reaction channel,  
\begin{equation}
\begin{aligned}
 (\mathcal{L}_B(\lambda))_{i,j} &\equiv \delta_{i,j+1}\left(\alpha_i^++\beta_i^+\e^{-\lambda}\right)+\delta_{i,j-1}\left(\alpha_i^- +\beta_i^-\e^\lambda\right)\\
 &-\delta_{i,j} \left(\alpha_i^+ + \alpha_i^- + \beta_i^+ + \beta_i^+ \right) ,
 \end{aligned}
\end{equation}
where $\alpha_i^\pm$ and $\beta_i^\pm$ are the transition rates for the $A$ and $B$ channels, respectively. A simple calculation shows that ${L}_A(\lambda)$ and ${L}_B(\lambda)$ are related by the following symmetry
\begin{equation}
 \mathcal{L}_B(\lambda) = \mathcal{A}^{-1} \mathcal{L}_A(\lambda) \mathcal{A},
 \label{eq:symm}
\end{equation}
where 
\begin{equation}
 \mathcal{A}_{i,j} = \delta_{i,j}\e^{\lambda j}.
\end{equation}
As \cref{eq:symm} describes a similarity transformation, $\mathcal{L}_A(\lambda)$ and $\mathcal{L}_B(\lambda)$ have the same eigenvalues. It then follows that $Z_A(t)$ and $Z_B(t)$ have the same scaled cumulant generating function, \cref{eq:SCGF}, as it is given by the largest eigenvalue of the tilted operator \cite{lebo99}. \\

\section{Calculation of the diffusion coefficient without relying on large deviation theory} \label{app:noLDT}
Here, we present a derivation of the diffusion coefficient without relying on large deviation theory.
We consider the two-state model introduced in \cref{sec:twostate}. For simplicity, we neglect contributions from jumps between fixed-points $\mathcal{B}_\pm$ and the diffusion around the fixed-points, see \ref{sec:twostate_sigma} for further explanations.  

Along a stochastic trajectory $n(t)$, the time-integrated current of $B$ molecules is given by
\begin{equation}
Z_B = \int_0^T \mathrm{d}t\left(J_-\delta_{n(t),-} + J_+ \delta_{n(t),+}\right).
\end{equation}
where $J_\pm$ is the flux of $B$ molecules in state $x_\pm$. The average flux is 
\begin{equation}
J_B = \lim_{T\rightarrow\infty} \frac{\avg{Z_B}}{T} = \sum_{n={-,+}} p_n J_n .
\end{equation}

To compute the diffusion coefficient, we will now consider a shifted system where the flux is $0$ in state $x_-$ and $(J_+-J_-)$ in state $x_+$. The shifted time-integrated current is
\begin{equation}
 \widetilde{Z}_B \equiv \int_0^T\,\mathrm{d}t J \delta_{n(t),+},
\end{equation}
and its associated flux
\begin{equation}
  \avg{\widetilde{Z}_B} = \avg{Z_B}-J_- =  T \left(J_+-J_-\right) p_+  .
\end{equation}
The second moment of $\widetilde{Z}_B$ is given by
\begin{equation}
 \avg{\widetilde{Z}_B^2} = \left(J_+-J_-\right)^2\int_0^T\mathrm{d}t\int_0^T\mathrm{d}t' \underbrace{\avg{\delta_{n(t),+}\delta_{n(t'),+}}}_{p(+,t;+,t')}
 \label{eq:zb2}
\end{equation}
where $p(+,t;+,t')$ is the joint probability to be in state $x_+$ at times $t$ and $t'$, in the steady-state it is equal to $p(+,t-t';+,0)$. We solve the two-state master equation and obtain
\begin{equation}
\begin{aligned}
 p(+,\tau;+,0) &= p_+\,p(+,\tau|+,0) \\
            &= p_+\left(p_+-(p_+-1)\e^{-\left(r_-+r_+\right) \tau} \right).
\end{aligned}
\label{eq:pjoint}
\end{equation}
By inserting this expression into \cref{eq:zb2}, we can compute the variance which does not depend on the shift. We obtain
\begin{widetext}
\begin{equation}
\begin{aligned}
  \avg{{Z}_B^2} - \avg{{Z}_B}^2 &= \avg{\widetilde{Z}_B^2} - \avg{\widetilde{Z}_B}^2 \\
  &= 2p_+\left(J_+-J_-\right)^2\int_0^T\mathrm{d}t\int_0^t\mathrm{d}\tau \left(p_+-(p_+-1)\e^{-\left(r_-+r_+\right) \tau}\right) - \avg{\widetilde{Z}_B}^2\\
 &=2p_+\left(J_+-J_-\right)^2\left(\frac{\left(1-p_+\right)}{\left(r_-+r_+\right)^2}\left(\e^{-\left(r_-+r_+\right) T}-1\right)+\frac{\left(1-p_+\right)T}{\left(r_-+r_+\right)}\right). \\
 \end{aligned}
\end{equation}

Finally, we get the diffusion coefficient
\begin{equation}
\begin{aligned}
 D_B &= \lim_{T\rightarrow\infty}\frac{\avg{{Z}_B^2} - \avg{{Z}_B}^2}{2T} = p_-p_+ \frac{\left(J_+-J_-\right)^2}{r_-+r_+}.
 \end{aligned}
\end{equation}
\end{widetext}

%

\end{document}